\documentclass{emulateapj}

\begin{document}
\title{Confinement of supernova
explosions in a collapsing cloud}
\author{M. M. Kasliwal, R.V.E. Lovelace, and J.R. Houck}
\affil{Department of Astronomy,
Cornell University, Ithaca, NY 14853;  MMK32@cornell.edu;
RVL1@cornell.edu;  JRH13@cornell.edu}

\begin{abstract}

        We analyze the
confining effect of cloud
collapse on an expanding
supernova shockfront.
        We  solve the differential
equation for the forces on the
shockfront due to ram pressure,
supernova energy, and gravity.
       We find that the expansion of
the  shockfront is slowed and in fact reversed
by the collapsing cloud.
      Including radiative losses and a potential
time lag between supernova explosion and cloud
collapse shows that the expansion is reversed at
smaller distances as compared to the non-radiative
case.
        We also consider the case of multiple
supernova explosions at the center
of a collapsing cloud.
      For instance, if we scale our
self-similar solution to a single supernova of energy
$10^{51}$ ergs occurring when a cloud of initial
density $10^{2}$ H/cm$^{3}$ has collapsed by $50\%$,
we find that the shockfront is confined to $\sim 15$ pc in
$\sim 1$ Myrs.
         Our calculations are pertinent to
the observed unusually
compact non-thermal radio emission
in blue compact dwarf galaxies (BCDs). More generally, we
demonstrate the potential of a collapsing cloud to confine
supernovae, thereby explaining how dwarf galaxies would exist
beyond their first generation of star formation.

\end{abstract}



\section{Introduction}

Blue Compact Dwarf galaxies (BCDs) are
observed to be experiencing
intense star formation in a spatially compact
region. The timescale for this star formation
episode of mass $\sim$ 10$^{8} M_{\sun}$ is
only a few million years.
         Massive stars have short lifetimes and thus
supernova (SN) explosions may occur concurrently in
regions of cloud collapse. It is quite probable that this
occurs in blue compact dwarf galaxies. In this paper,
we  study the confining effect of cloud collapse on
an expanding supernova shockfront.

        The forces acting on the shockfront
include those due to supernova energy,
ram pressure and gravity.
When a supernova shockfront expands into a stationary cloud 
(Sedov 1946, Taylor 1950, Bisnovatyi-Kogan \& Silich 1995), 
the accreted mass does not contribute to a change in the 
momentum flux. In a region of cloud collapse, not only
is mass accreted at a faster rate, it also contributes
to a change in the momentum flux. This additional ram pressure
leads to the confinement of a supernova shockfront in a collapsing
cloud. The potential of ram pressure to confine stellar winds of 
a noncentral OB star to generate a steady-flow situation has been shown 
(Dopita 1981). We demonstrate that ram pressure can also confine expanding 
supernova shockfronts with pressures as high as $\sim$ 10$^{-9}$ dynes/cm$^{2}$ 
as compared to stellar wind pressure which is $\sim$ 10$^{-12}$ dynes/cm$^{2}$
(Dopita et al 1981). Our self-similar solution describes the 
time evolution of a central supernova explosion in a collapsing cloud.

We start with a discussion of the
gravitational collapse of a cloud,
and then obtain the equation for
the expansion of a SN shock in
the cloud. We first neglect
radiative losses and then
include them. Next, we consider
a time lag between when the supernova explodes
and when the cloud begins to collapse.
        Finally, we consider the
case of multiple SN explosions
occurring at the center
of a two-component collapsing cloud.
The inner component has higher
density and collapses rapidly
to produce the central stellar
population. A few million years
later, the resulting supernova
shockfronts collide with the
infalling, less dense,
outer component of the cloud.

\section{Model}

We first consider a simple model
for the free-fall gravitational collapse
of a cloud with negligible pressure.
The gravitational force on a
differential spherical shell  with
initial radius $R_0$ and radius  $R(t)$, is
\begin{equation}
\label{CC1}
dM(R)~\frac{d^2R}{dt^2}=
-\frac{GM(R_0)~dM(R)}{R^2}~,
\end{equation}
where G is the universal gravitational
constant and $M(R_0)$ is the initial
mass enclosed by this shell.
$
M(R_0)=4\pi R_0^3\rho_0/3,
$
where $\rho_0$ is the initial
density of the cloud which is
assumed to be uniform.
        Solving this equation,
assuming
the initial velocity is zero,
gives
\begin{equation}
\label{VcR}
{dR \over dt}=
-\left[2GM(R_0)\left(\frac{1}{R}-
\frac{1}{R_0}\right)\right]^{1/2}~.
\end{equation}
       The free-fall time of the cloud is
\begin{eqnarray}
\label{tff}
{t_{ff}}&=&\int_{0}^{R_0} \frac{dR}{-dR/dt}
=\left(\frac {3\pi}{32G\rho_0}\right)^{1/2}
\nonumber\\
&\approx& {5.2\times 10^6 {\rm yr}}~{(n_{2})^{-1/2}}~,
\end{eqnarray}
where $n_{2}$ is density in units of
10$^2$ H/cm$^{3}$.
The free-fall time is
independent of the initial radius. Thus,
the collapse is homologous
as shown in Figure 1.

       The velocity of the shell hitting the shock
front contributes to the ram  pressure that
slows down the  front.
        We normalize the time
and radius as $\tau={t}/{t_{ff}}$
and $\overline{R}={R}/{R_0}$.
Hence we obtain
\begin{equation}
\label{tau}
\tau=\frac{2}{\pi}
\int_{\overline{R}}^{1}  dr^\prime
\left(\frac{r^\prime}{1-r^\prime}\right)^{1/2}~.
\end{equation}
This  integral  gives
a cumbersome expression for
$\overline{R}(\tau)$.
        Therefore, we
approximate the dependence
as
\begin{equation}
\label{ftau}
\overline{R} \approx (1-\tau^2)^{1/2}\equiv f(\tau)~.
\end{equation}
       Thus the velocity of the cloud at the
shockfront radius, $R_{sh}$, is
\begin{equation}
\label{Vc}
{V_{cl}(\tau)=
-\frac{\pi R_{sh}[1-f(\tau)]^{1/2}}
{2t_{ff}{[f(\tau)]^{3/2}}}}~.
\end{equation}

      Accretion of mass just outside
the shock front acts to
increase in density.
Mass conservation gives
\begin{equation}
\label{delM}
\Delta M =
4\pi R_0^2 \Delta R_0\rho_0 =
4\pi R^2 \Delta R\rho(t) ~.
\end{equation}
Using $R=R_{0}f(\tau)$ and $\Delta R=
\Delta R_{0}f(\tau)$ gives
\begin{equation}
\label{rho}
\rho(\tau)=\frac{\rho_0}{[f(\tau)]^3}~.
\end{equation}

There are three forces experienced  by the shock front:
(i) The energy released by the supernova
explosion $E_{SN}$ inside the shell  gives
an outward directed pressure
$p=(\gamma - 1)E_{SN}/(\frac{4}{3}\pi R_{sh}^3)$
or force $4\pi R_{sh}^2 p$. We use $\gamma = 5/3$.
(ii) The ram pressure of the cloud shell
hitting the shock front results in
an inward directed force (or pressure).
When the direction of motion of the
external medium and the
shockfront are the same, the ram pressure acts only if
the speed of the medium is greater than that
of the shock front.
        This is accounted for by a
      Heaviside function, $H$, of the difference between
the shockfront velocity and cloud velocity
(with $H(>0)=1$ and $H(<0)=0$).
(iii) The gravitational force is of course inward.

The ambient pressure of the interstellar medium
is $\sim$ 10$^{-10}$ dynes/cm$^{2}$ 
for a density $\sim$ 100 ${\rm H/cm}^{3}$ and temperature $\sim$ 10$^{4}$ K.
For a typical speed $\sim$ 100 km/s, the ram pressure is $\sim$
10$^{-8}$ dynes/cm$^{2}$. For this reason, we neglect the ambient 
pressure of the interstellar medium.

The rate at which the SN shockfront accretes mass is proportional
to the relative velocity between the cloud and the shockfront. Thus,

\begin{equation}
\label{dMtau}
\frac{dM_{sh}}{dt} = 4\pi{R_{sh}}^2~\rho~U~,
\end{equation}
where
\begin{equation}
U \equiv {{dR_{sh}}\over{dt}}-V_{cl}~.
\end{equation}

The accreted mass initially moved
at the cloud velocity. It thus
contributes to a net change in the momentum flux.
For the case where
the SN explosion and cloud collapse
begin at the same time we have
\begin{equation}
\label{DEQ}
\frac{d}{dt}\left(M_{sh}\frac{dR_{sh}}{dt}\right) =
4\pi R_{sh}^2\left[\frac{E_{SN}}{2 \pi R_{sh}^3}
+ \rho H(U)~U~V_{cl} \right]
-\frac{GM_{sh}^2}{2R_{sh}^2}~.
\end{equation}

We normalize time by the free-fall time
and the shockfront radius by the
Sedov-Taylor  radius at the free-fall
time (within a factor of 1.24); that
is,
$\overline{R}_{sh}={R_{sh}}/{R_{ST}(t_{ff})}$, where
\begin{equation}
R_{ST}(t_{ff})\equiv ((\gamma - 1)E_{SN}t_{ff}^2/\rho_0)^{1/5}
\approx 52~{\rm pc}~
(E_{51})^{1/5}
(n_{2})^{-2/5}~.
\end{equation}
We normalize mass by
$\overline{M}_{sh}={M_{sh}}
/[4 \pi\rho_0 R_{ST}^3(t_{ff})/3]$.

        We define a `momentum' variable
\begin{equation}
\label{P}
{\cal P} \equiv \overline{M}_{sh}
\frac{d\overline{R}_{sh}}{d\tau}~.
\end{equation}

Rewriting equation (\ref{dMtau}) and equation (\ref{DEQ})
in terms of the normalized parameters gives

\begin{equation}
\frac{d\overline{M}_{sh}}{d\tau} =
\frac{3{\overline{R}_{sh}}^{2}}{f^3}\left(\frac{d\overline{R}_{sh}}{d\tau}+\frac{\pi\overline{R}_{sh}({1-f})^{1/2}}{2f^{3/2}}\right)~.
\end{equation}

\begin{equation}
\frac{d{\cal P}}{d\tau} =
\frac{{9}}{4\pi{\overline{R}_{sh}}}
-\frac{3\pi{\overline{R}_{sh}}^{3}~
({1-f})^{1/2}~H(\overline{U})~
\overline{U}}{2~f^{9/2}}
-\frac{\pi^2\overline{M}_{sh}^{2}}{16{\overline{R}_{sh}}^2}~,
\end{equation}
where
\begin{equation}
\label{Ubar}
\overline{U}=
\frac{\cal P}{\overline{M}_{sh}}
+\frac{{\pi\overline{R}_{sh}}
({1-f})^{1/2}}{2{f}^{3/2}}~.
\end{equation}
Thus, we have a system of three
non-linear coupled first order equations.
       For the initial conditions
we assume  at an early time, $R_{sh}$
and ${dR_{sh}}/{d\tau}$
are given by the standard
Sedov-Taylor solution for a stationary
cloud. We use the initial condition
that at $\tau=10^{-6}$,
$\overline{R}_{sh}=0.0049$,
$\overline{M}_{sh}=1.2 \times {10}^{-7}$
and $\cal{P}$$ = 0.00024$.
        Figure 2 shows $\overline{R}_{sh}(\tau)$
and the corresponding Sedov-Taylor solution.
        The infalling cloud  at first slows
the shock expansion and later reverses
its motion.

\section{Influence of Radiative Losses}

In the above analysis  we assumed
that the energy within the
SN shock was constant, but
this is valid only for relatively
short times.
       We next take into account the
radiative losses using the
pressure driven snowplow
model of Cioffi,
McKee,  and Bertschinger (1988).
        We find that the supernova
is confined to a much smaller
region in a shorter time.
      We use the mean pressure
$\overline{p}_{rl}$ including
radiative losses derived by
Cioffi et al. (1988)
which can be written as
     \begin{equation}
\label{Prad}
\overline{p}_{rl}=
\frac{\alpha~E_{SN}^{5/3}}{R_{sh}^5~t^{4/9}~
\rho^{10/9}~\eta_M^{4/7}}~.
\end{equation}
where magnitude of $\alpha$ is $2.3 \times 10^{-17}$ and $\eta_M$ is
metallicity in units of solar metallicity.
Thus $\overline{p}_{rl}$ replaces
our earlier pressure $p=E_{SN}/(2\pi R_{sh}^3)$.
Next, we assume that the
metallicity is solar metallicity
and we use equation (\ref{rho})
for density as a function of time.
The differential equation including
the radiative losses is
\begin{equation}
\label{DEQrad}
\frac{d}{dt}\left(M_{sh}\frac{dR_{sh}}{dt}\right) =
4\pi~R_{sh}^2\left(\overline{p}_{rl}
- \rho H(U)~U~V_{cl} \right)
-\frac{GM_{sh}^2}{2R_{sh}^2}~.
\end{equation}
We continue to normalize time
by the free-fall time.
       However, we now normalize the radius by
\begin{eqnarray}
\label{Rrad}
{R_{rl}}&=&\left({\frac{3\alpha~E_{SN}^{5/3}~
t_{ff}^{14/9}}{\rho_{0}^{19/9}}}\right)^{1/7}
\nonumber\\
&\approx& {11.8 {\rm pc}~(E_{51})^{5/21}~(n_{2})^{-26/63}}~.
\end{eqnarray}

We again obtain a self-similar
differential equation and we
solve it using the initial
condition that the shockfront follows
the Sedov solution at $\tau=10^{-6}$.
We use this initial condition
because   the
transition from the Sedov solution to
the pressure driven snowplow radiative
shockfront occurs at
\begin{equation}
t_{PDS}\approx 1.33 \times 10^4~{\rm yr}~
\frac{E_{51}^{3/14}}{n_0^{4/7} },
\end{equation}
(Cioffi et al. 1988).
For $E=10^{51}$ ergs and $n_0=10^2 {\rm H/cm}^{3}$,
$t_{PDS} \approx 1700 $ yr $ \sim t_{ff}/3000$.

The SN explosion may occur at
a time $\Delta t$ after the cloud collapse has begun.
         The velocity and density of the shells
hitting the shock front is then larger.
        We take the zero of time $t$
or $\tau=t/t_{ff}$ to be the
time when the supernova explodes so that
only equation (\ref{ftau}) needs to be
modified and replaced by
$f(\tau)=\overline{R}=
[1-(\tau+\Delta\tau)^2]^{1/2}
$, where $\Delta \tau =\Delta t/t_{ff}$ is
termed the ``lag.''

Figure 3 shows sample results for different lags which shows that the
shock expansion is more strongly decelerated the longer the lag. 
For no time lag, the shockfront is confined to $\overline{R}_{sh}$
= 1.99 and $\tau$ = 0.51. In comparison, for 50\% time lag, 
the shockfront is confined to $\overline{R}_{sh}$ = 1.33 and 
$\tau$ = 0.23 and for 80\% time lag, the shockfront is 
confined to $\overline{R}_{sh}$ = 0.75 and $\tau$ = 0.09.

\section{Multiple Supernova Explosions}

We now consider a more complete physical
picture. Suppose that there is a
localized clump of density $10^4$ H/cm$^3$
in a cloud of density $10^2$ H/cm$^3$.
     Suppose that the ambient outer cloud
and the inner clump begin to free-fall.
      The clump collapses in $\sim 0.5$ Myrs and forms
supernovae in another $2-3.5$ Myrs (lifetime of $30-24
M_{\sun}$ star). The free-fall time
of the cloud is $\sim 5$ Myrs and thus, the time lag
between when the supernova explodes and when the cloud
begins to collapse  ranges from $0.5 t_{ff}$ to
$0.8 t_{ff}$.

When the localized clump collapses, the initial mass
function is  such  that many low mass (lm) stars
form at the same time as a few high mass
stars.
      The distribution is dN/dM $\propto$ M$^{-2.35}$ (Salpeter 1955).
If we take
the high mass (hm)  threshold at $> 20 M_{\sun}$,
then $M_{lm} \sim
200N_{hm}M_{\sun}$.
     The mass of the low mass stars contributes
to the gravitational  force in addition to the forces described
in equation(\ref{DEQrad}).  The right-hand side of this
equation becomes
\begin{equation}
4\pi R_{sh}^2\left[\overline{p}_{rl}
- \rho H(U)~U~V_{cl} \right]
-\frac{GM_{sh}^2}{2R_{sh}^2}
-\frac{GM_{sh}M_{lm}}{R_{sh}^2}~.
\end{equation}
      We find however that the
gravitational force contribution
of the low mass stars does not significantly
alter the trajectory of the SN shockfront.

\section{Discussion}

Regarding the stability of the shockfront,
notice that the
    Rayleigh-Taylor
instability occurs in a
stratified region
where the effective gravity
points in the direction of the
less dense medium.
    In the case of an expanding supernova
shockfront, the effective
gravity in the frame of the shockfront
is the shockfront acceleration.
Our results for the trajectory are
all concave down implying that
effective gravity points radially outward
and thus, away from the relatively rarer medium.
Hence, the shockfront is Rayleigh-Taylor stable.

We conclude that ram pressure
from a collapsing cloud reduces
the shockfront velocity of one
radiative supernova explosion (occurring
in a localized region when the ambient
cloud has collapsed by 50\%) to zero at a
normalized radius of $1.33$ and a
normalized time of $0.23$.
      These
fractions scale by the energy of
the supernova and the density of
the interstellar medium.

Observations of nonthermal radio emission in galaxies 
suggest whether or not the supernovae are confined. For
example, a 22 GHz map of M82 extending over a few 
hundred parsecs is estimated to have 40 unconfined supernovae 
(Golla et al 1996).  An 18 cm nonthermal radio emission map of VIIZw19 
extending over 310 pc is estimated to have 2500 unconfined supernovae remnants
(Beck. et al 2004). Two examples of apparently confined emission, SBS-0335 and Henize 2-10, are described below.

In a model of the
nonthermal radio emission from SBS-0335 the emission is
confined to $17$ pc (Hunt et al. 2004)
within a $520$ pc region
(Thuan et al. 1997) where star
formation occurs in six super-star
clusters with ages $\leq 25$ Myrs.
  If a supernova explosion of
energy $10^{51}$ ergs occurs
after a cloud of initial density 10$^2$ H/cm$^{3}$
has collapsed by 50\%,it is confined to
$\sim 15$ pc in $\sim 1$ Myr.
If this were a supernova exploding in a stationary cloud, it would
expand beyond  $\sim 20$ pc in as little as 0.4 Myrs. This
would destroy the cloud and the case of a young supernova
remnant in a stationary cloud is considerably less probable
than a supernova confined in a collapsing cloud.

Very Large Array imaging of another BCD, Henize 2-10,
indicates a $<8$ pc region of $1$ mJy radio
sources in the central $5^{''}$ starburst region.
Henize 2-10 has HII regions
of sizes between $3$ pc and $8$ pc and densities
between $1500$ and $5000$ H/cm$^{3}$
(Kobulnicky, Johnson 1999).
If we hypothesize that a cloud of an average initial density of
$3000$ H/cm$^3$ has collapsed by $50\%$ when $10$ supernovae
explode, we find that
the supernovae are confined to
$\sim 7$ pc in $\sim 0.2$ Myr.
Thus, cloud collapse successfully confines
supernova explosions and can  account for observed
compact nonthermal radio emission. This simple model
can help understand continuous or second/third generation
star formation since it suggests why the cloud is not
devastated by first generation supernovae.

We  thank David Chernoff, Ruizhen Tan,
and Setu Mohta for their valuable comments.
    Also, we thank the referee for helpful
suggestions.  Support
for this work was provided by NASA through contract
1257184 issued by JPL/Caltech. The work of RVEL was
supported in part by NASA grants NAG 5-13060,
NAG 5-13220, and NSF grant AST-0307817.

\clearpage
\begin{figure*}[h]
\epsscale{1}
\plotone{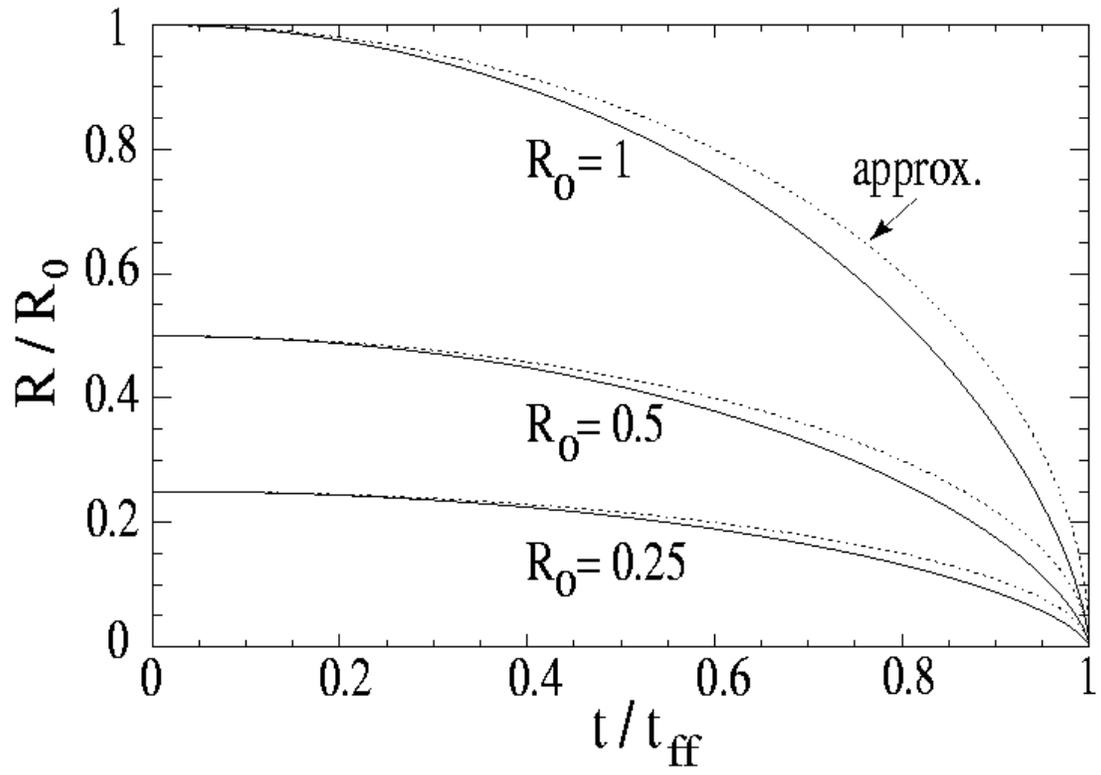}
\figcaption{The collapse  of cloud
shells with different initial radii
are shown as solid lines.
The radius is normalized by the
initial radius of the outermost cloud
shell and the time is normalized by
the free-fall time.
       The dotted line
indicates the approximation of
the trajectory used here. The
approximation has a maximum deviation
of 5.3\% and a mean deviation of
2.6\%.}
\end{figure*}

\clearpage
\begin{figure*}[h]
\plotone{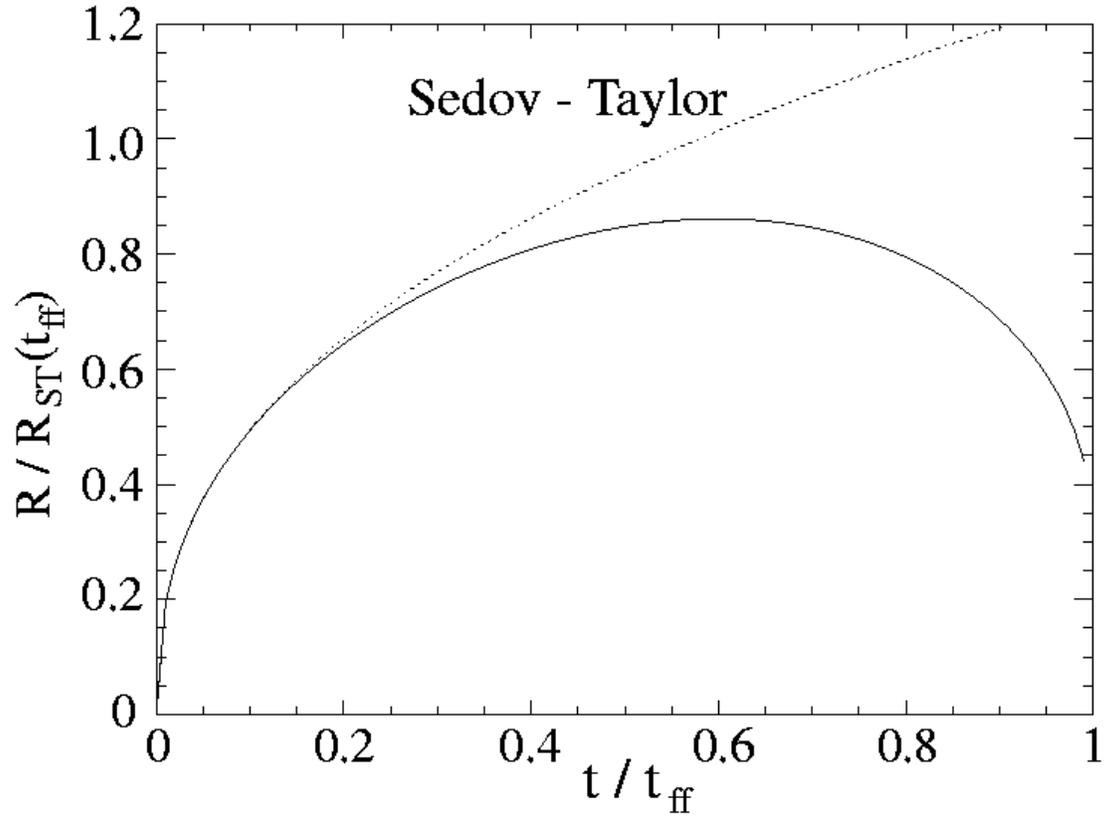}

\epsscale{1}
\figcaption{Shock wave expansion $R_{sh}(\tau)$
in a collapsing cloud and
the standard Sedov-Taylor
expansion in a stationary medium.
       Time $\tau$ is normalized by the
free-fall time and the radius is
normalized by the Sedov-Taylor radius
      at the free-fall time.
        The shock expansion is slowed
down and subequently turned around by the collapsing cloud.
}
\end{figure*}

\clearpage
\begin{figure*}[h]
\plotone{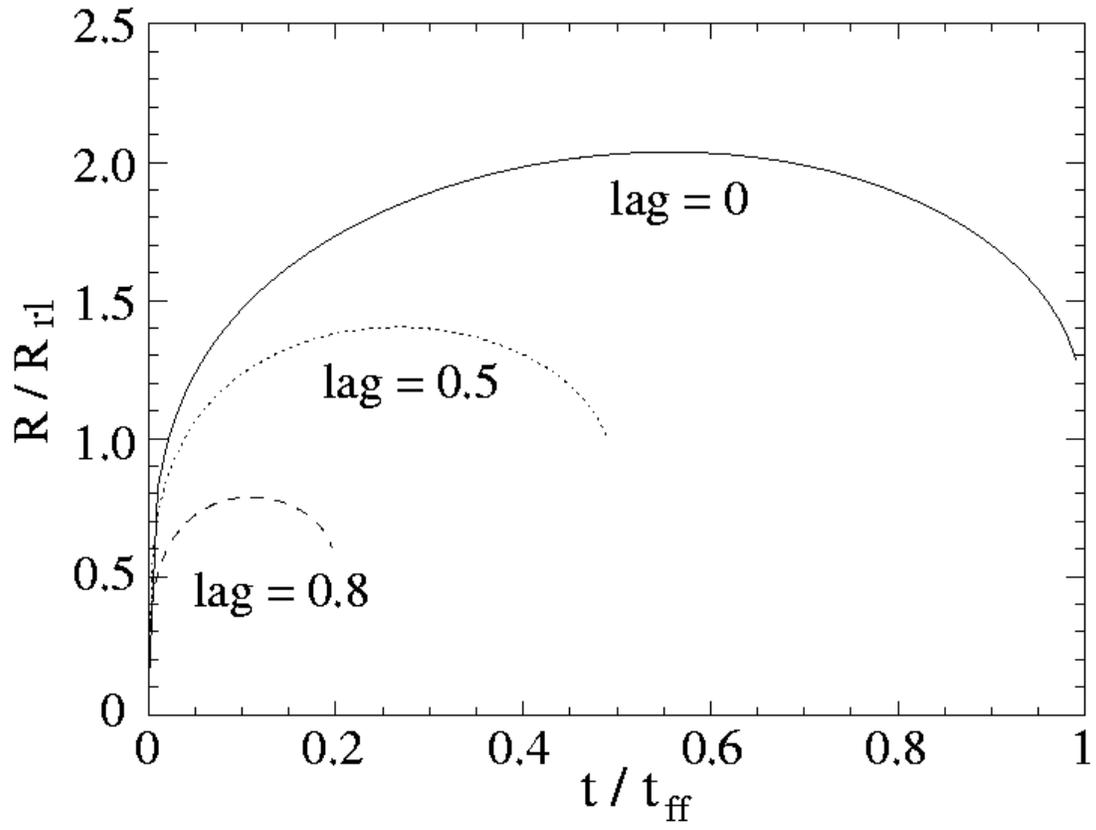}
\epsscale{1}
\figcaption{Shockfront radii
including influence of the radiative losses.
We also show the shockfront radius for a time
lag of 50\% and 80\% of the free-fall
time between supernova explosion and cloud collapse.
Time zero is taken as the instant the supernova
exploded.}
\end{figure*}

\clearpage
\begin{figure*}[h]
\plotone{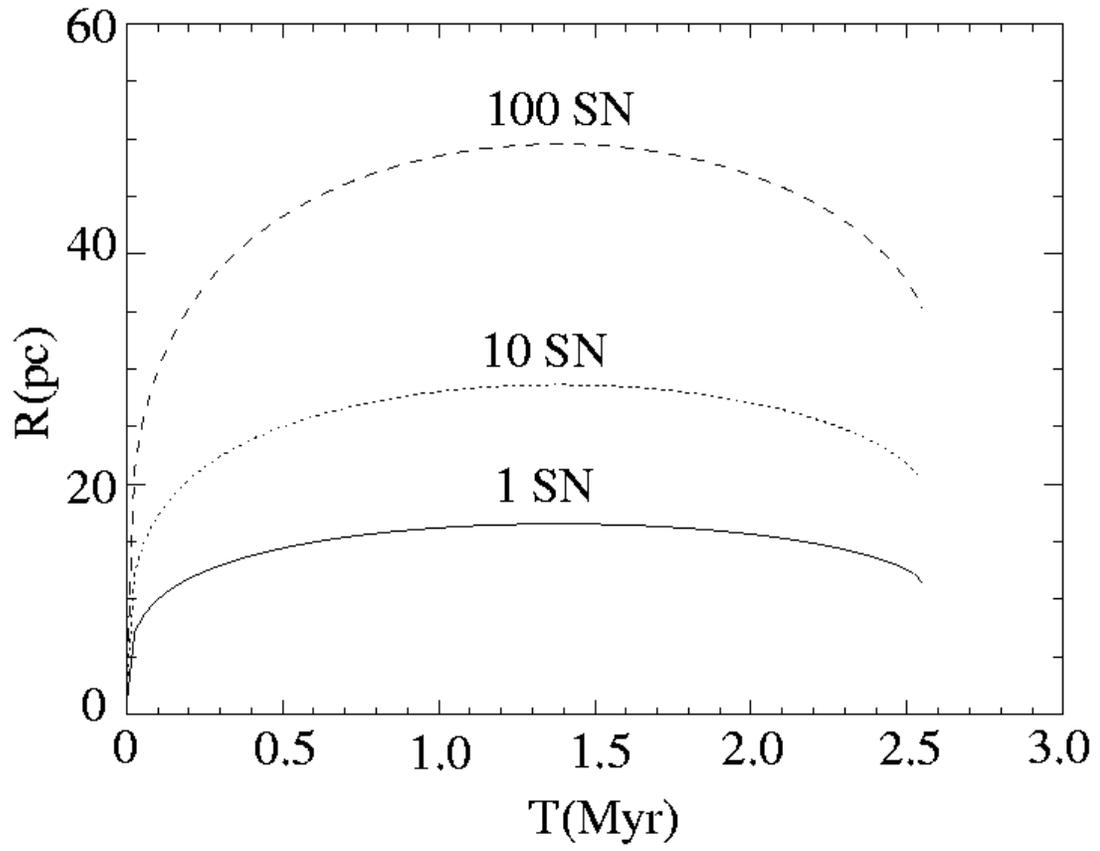}
\epsscale{1}
\figcaption{Evolution
of the shockfront radius for
single and multiple SN explosions.}
\end{figure*}

\end{document}